\magnification=1200
\hsize=15.6 truecm
\vsize=23.0 truecm

\centerline{\bf MECHANICS OF COSMIC RINGS}
\vskip 1 cm
\centerline{\bf Brandon Carter}
\vskip 0.4 cm
\centerline{\it D\' epartement d'Astrophysique Relativiste et de
Cosmologie, C.N.R.S,}
\centerline{\it Observatoire de Paris, 92 195 Meudon, France.}
\vskip 0.4 cm
\centerline{January, 1990.}
\vskip 1.0 cm

\noindent
{\bf Abstract.} 

In a flat background, simple non-conducting string loops have no strictly
stationary equilibrium states, but for cosmic string loops of 
superconducting kind such ``vorton'' states will exist, with rotating 
circular configurations appropriately describable as ``cosmic rings'' 
(which may conceivably be a significant contributor to the material 
content of the universe) whose equilibrium is obtained for zero angular 
velocity of retrograde characteristics. Such a ``cosmic ring'' is 
characterised by its mass $M$ and angular momentum $J$ say, 
corresponding at a microscopic level to two independent quantum numbers, 
namely charge number $C$ and a topological phase number $N$ whose 
product determines the angular momentum, $J=CN$, while their ratio 
determines the local intrinsic state of the string, which may be 
qualitatively classified as being of ``electric'' or ``magnetic'' type 
depending on whether $C/N$ is greater or less than a critical value 
dependent on the underlying theoretical model.

\vskip 1.0 cm

\bigskip\noindent
{\bf 1. Introduction}
\medskip

The purose of this work is to set up the elements of the theory of what 
we shall refer to as ``cosmic rings'' in a framework [1,2] that is 
general in the sense of being independent of the precise details of the 
underlying quantum field theory that may be postulated to give rise to 
the cosmic string model under consideration. We use the term ``cosmic 
ring'' to desribe a circular equilibrium state of a local cosmic string 
loop in a flat background. Our present discussion is based on the 
supposition that gravitational and electromagnetic self-interaction are 
sufficiently weak to be neglected. It is of course well known that no 
such strictly stationary equilibrium states can exisst for a simple 
Goto-Nambu type cosmic string loop. However, for a cosmic string loop of 
the superconducting kind [3] the possibility of centrifugally suppported
circular equilibrium states has been clearly recognised by Davis and
Shellard [4], who have drawn attention to the interest of such rings
as examples of ``vorton'' type semi-topological solitons, which may
conceivably have been produced in considerable numbers in the early
universe (perhaps even so copiously as to pose a hidden matter problem
analogous to that of monopoles).

Davis and Shellard have plotted numerical results [4] for a number of
relevant quantities for such ``cosmic ring'' configurations on the
basis of a particular field theoretical model of the kind commonly
postulated as an underlying framework for the theory of superconducting
strings as introduced by Witten [3]. Such models [5,6] can be considered
at a macroscopic level as particular examples within the broad category
of electromagnetic string models describable by the recently introduced
covariant formalism [1,2] which provides a convenient foundation for a
general derivation of the essential elements of the theory of cosmic ring
states, making it possible to extend the work of Davis and Shellard,
complementing their numerical results by general analytic results 
applicable to the entire class of conceivably relevant models.

In the discussion of Davis and Shellard [4], terms such as ``static''
and ``stable'' were used rather loosely for what in stricter terminology
would be characterised as ``stationary'' and ``equilibrium''. The question
of strict global {\it stability} of the equilibrium states under
consideration has not yet been dealt with rigorously, but the necessity
at least of the standard local stability conditions [2] is evident, while
a further condition that may be conjectured to be sufficient is derived
below. The stationary rotating states we are considering here are not to
be confused with the strictly {\it static} states whose conceivable 
existence has recently been the subject of discussion by Hindmarsh, Turok
and coworkers [7,8], but whose existence is only possible for the 
restricted subclass of string models that allow vanishing tension, $T=0$,
which is necessary for equilibrium in the strictly static case. It is to
be mentioned that the term ``cosmic spring'' has sometimes been employed
in this latter context, entailing the implicit suggestion that such states
might be locally stable with respect to small oscillations betwwen states
of small positive tension and compressed states of small negative tension.
However, the introduction of the term ``spring'' in the present context is
misleading except for loops that are so short compared with the microscopic
thickness of the underlying quantum vortex that description as a ``string''
is no longer appropriate. For any loop sufficiently long compared with its
microscopic thickness for description as a ``string'' to be reasonable,
states of negative tension will inevitably be locally unstable [2] so that
the intermediate ``relaxed'' states of zero tension, if allowed at all by
the underlying field theory, would be at most marginally stable, and as 
such may appropriately be described not as ``springs'' but just as   
``loose strings''. In so far as it has been suggested that such 
{\it static} ``loose string'' states may conceivably be cosmologically
important, the same applies a fortiori to the ``ring'' states under
consideration in the present work, since the latter can exist for {\it any} 
superconducting string model, not just for a restricted subclass.
It is to be noticed that their properties are not unsimilar to those 
recently postulated for the charges ultra-massive particles (``chumps'')
that have recently been proposed as dark matter candidates [9].

In order to be effectively thin, so as to be appropriately qualified at
a macroscopic level as a ``string'' (and hence for its axisymmetric
``vorton'' equilibrium states to be describable as ``rings'' in the
technical sense used here), a vortex defect of the vacuum for a field
theory with spontaneously broken symmetry should be of ``local'' rather
than ``global type''. For the superconducting string models in question
[1], the macroscopic action (as obtained after integration over the
microscopic cross section of the local vacuum defect region) has the
form of an integral over a 2-dimensional world sheet (specifying the
mean, macroscopic, localisation of the string) of an effective lagrangian
function $L$ that depends only on the pseudo-metric norm $w$ of the
gauge covariant derivative within the world sheet of a scalar phase
field $\phi$. it will be convenient to introduca an important auxiliary
function $K$ constructed from the lagrangian by differentation according
to the prescription
$$ K=-2 {\rm d} L/ {\rm d} w\, ,\hskip 1 cm w=|D\phi|^2\, \eqno{(1.1)}$$
and it will also be useful to construct from $L$ the corresponding
``dual'' lagrangian function $\tilde L$ in which the roles of space and
time are interchanges [1], according to the prescription
$$\tilde L=L+ w K.  \eqno{(1.2)}$$

With our chosen sign convention for the spacetime pseudo-metric, 
a positive value of $w$ caracterises a ``magnetic'' string regime [1]
for which the gauge covariant derivative is space-like, so that the string
has a preferred rest frame density $U$ and a tension $T$ given by
$U=-L$, and $T=-\tilde L$, with an associated stream number density $\nu$
and an effective mass $\mu$ given respectivelu by $\nu^2=w$ and
$\nu=K\mu$. On the other hand a negative value of $w$ characterises an
``electric'' regime for which the covariant phase derivative is timelike,
so that the corresponding expressions are $T=-L$ and $U=-\tilde L$ for
the tension and energy density, with $\mu^2=-w$ and $\nu= K\mu$ for the
effective mass density and number density.

In both the ``magnetic'' and ``electric'' regimes the string 2-surface
stress-energy tensor is given in terms of time-like and space-like unit 
vectors $u^\mu$ and $v^\mu$ of the intrinsically preferred orthonormal
basis in the string 2-surface by an expression of the form
$$ T^{\mu\nu}=(U-T)u^\mu u^\nu -T \,\eta^{\mu\nu}\, ,\hskip 1 cm
\eta^\mu\nu=-u^\mu u^\nu+v^\mu v^\nu\, ,\eqno{(1.3)}$$
where $\eta^{\mu\nu}$ is the ``fundamental tensor'' of the world sheet 
[1]. The expression (3) is thus valid everwhere except perhaps at a 
``transluminal boundary'' locus on which the string passes through a
degenerate intermediate regime characterised by $T=U=-L=-\tilde L$,
where the stress-energy tensor will take the form $T^{\mu\nu}=
l^\mu l^\nu-T\,\eta^{\mu\nu}$, for some appropriately normalised null
eigenvectot $l^\mu$, the unit vectors $u^\mu$ and $v^\mu$ being
indeterminate.

It is to be noticed that the physical relevance of the foregoing 
formalism (and hence of the results to be described below) is not
restricted to strings of ``cosmic'' origin but applies also to the elastic
string models that are appropriate for familiar terrestrial applications
to ordinary ropes and wires (always in the limit when they can be 
considered to be sufficiently thin compared with the lengths involved), 
though in such cases the relevant scalar variable $\phi$ turns up
merely as an abstract auxiliary (Clebsch type) potential, lacking the
interpretation as a topologically periodic quantum phase that applies
in the cosmic case.

\bigskip\noindent
{\bf 2. Circular string loops}
\medskip

Our purpose here is to consider the simplest case of stationary circular
configurations characterised by a radius, $r$ say, and an angular
velocity ${\mit\Omega}$ say, of the preferred rest frame vector $u^\mu$,
subject to the subluminal rotation conditions $0<{\mit\Omega}^2
<r^{-2}$ (the limits ${\mit\Omega}^2=0$ and ${\mit\Omega}^2=r^{-2}$
correspond to the static limit and the transluminal limit respectively).
The dually associated (superluminal rotation) angular velocity that
is associated with the preferred space-like vector $v^\mu$ will be
given by
$$ \tilde{\mit\Omega}=1/r^2{\mit\Omega}\, .\eqno{(2.1)}$$
Under the foregoing conditions, it can be seen that, with respect to an
appropriately oriented local stationary background frame, the components
of the intrinsically preferred string 2-surface frame vectors will be
given by expressions of the form $u^{_0}=v^{_1}=1/\sqrt{1-r^2
{\mit\Omega}^2}$ and $u^{_1}=v^{_0}=r{\mit\Omega}/\sqrt{1-r^2
{\mit\Omega}^2}$. The ensuing expressions for the stationary bckground
components of the stress-energy tensor, namely
$$ T^{_{00}}={{U-r^2{\mit\Omega}^2 T}\over{1-r^2{\mit\Omega}^2}}\, ,
\hskip 1 cm T^{_{01}}={{r{\mit\Omega}(U-T)}\over{1-r^2{\mit\Omega}^2}}
\, ,\hskip 1 cm T^{_{11}}={{r^2{\mit\Omega}^2U-T}
\over{1-r^2{\mit\Omega}^2}}\, , \eqno{(2.2)}$$
enable us to evaluate the relevant mass function as given by
$$M=2\pi r\,  T^{_{00}} \eqno{(2.3)}$$
and the corresponding angular momentum
$$ J=2\pi r^2 T^{_{01}}\, .\eqno{(2.4)}$$
It is apparent from these expressions that the angular momentum and
mass functions are related by the dually alternative expressions
$$M={\mit\Omega} J+ 2\pi r\, U=\tilde {\mit\Omega} J+2\pi r \, T
\, .\eqno{(2.5)}$$ 
Since we are supposing that the coupling is sufficiently weak for 
external forces on the string to be negligible we may choose to work in a 
gauge such that the Maxwellian connection form $A_\mu$ vanishes so that 
the components of the covariant derivative are simply $D^{_0}\phi=\omega$
and $D^{_1}\phi=k$ giving the pseudo-norm $w$ that plays the role of the 
independent intrinsic state variable in the form
$$w=k^2-w^2\, ,\eqno{(2.6)}$$
where $\omega$ is the frequency and $k$ the wavenumber of the phase field.
Under these conditions the electric current vector $I^\mu$ will have
components given with respect to the stationary background frame by
$$I^{_0}= e\, K\, \omega\, ,\hskip 1 cm I^{_1}= e\, K\, k\, ,
\eqno {(2.7)}$$
where $e$ is the charge coupling constant. On the understanding that
in the present section we are using units such as to give unit value
not only for the speed of light $c$ (as has been asumed throughout) but 
also for the Dirac-Planck constant $\hbar$, the corresponding magnetic 
dipole moment, $D$ say, and the electric monopole moment, $Q$ say, for 
the ring will be given by
$$D=\pi r^2 I^{_1}=r\, Q\, k/2\omega \eqno{(2.8)}$$
and
$$Q= 2\pi r I^{_0}=e\, C\, ,\eqno{(2.9)}$$
where the conserved charge number $C$ is an integer
valued quantity given by
$$C=2\pi r\, K\,\omega\, ,\eqno{(2.10)}$$
the qualitative physical interpretation of this expression being 
dependent on whether we are dealing with the ``magnetic'' case  for 
which $K=\mu/\nu$ or the ``electric'' case for which $K=\nu/\mu$. The
independent topologically conserved quantum number representing the
winding number of the phae round the ring can be seen to be expressible 
even more simply by 
$$ N= r\, k\, .\eqno{(2.11)}$$
Using the fact that the phase velocity $\omega/k$ directly determines the
angular velocity by the proportionality relation $\omega/k=r{\mit\Omega}$
in the ``magnetic'' cae, i.e. for $w>0$, and that it analogously 
determines the dual angular velocity  $\omega/k=r\tilde{\mit\Omega}$ in
the ``electric'' case, i.e. for $w<0$, we see from the expressions above 
that the angular momentum quantum number will be given in either case
simply as the product of the charge number and the phase winding number:
we shall always have
$$ J=C\, N\, .\eqno{(2.12)}$$

\bigskip\noindent
{\bf 3. The standing wave condition for equilibrium}
\medskip

The configurations that we are considering have up to this stage not
only been dependent on the two independent parameters, $\omega$ and $k$ 
say (which together determine the intrinsic state parameter $w$) but are
also dependent on a third independent overall scale parameter which may
conveniently be taken to be the radius $r$. However, the number of
independent parameters reduces from three to two when one takes 
account of the condition for centrifugal equilibrium, which is equivalent
to the condition that the mass function be stationary with respect to
variations subject to the constraints that the separate quantum numbers
$C$ and $N$ (and hence also $J$) are held constant. It is to be remarked
that these constraints are such as to ensure automatically the
preservation of the qualitative ``magnetic'' or ``electric'' character
of the ring which cannot be affected by any continuous variation that
preserves the ratio $C/N=2\pi\, K\,\omega/k$, ``magnetic'' and 
``electric'' regimes being separated by the critical values 
$C/N=\pm 2\pi\, K_0$, where $K_0$ is the value of the function $K$ where 
its argument vanishes, i.e. for $w=0$, which would be attained if the 
phase velocity passed through the speed of light, $\omega/k=\pm 1$. (The 
qualitatively intermediate case of what might be termed a 
``transluminal'' ring would be envisageable as a physically attainable 
intermediate state, within the framework of the simple string model under 
consideration, if the underlying theory were such as to provide a 
lagrangian function for which $2\pi\,K_0$ happened to be a rational 
ratio, so that the positive and negative critical values of $C/N$ could 
be realised for integer values of $C$ and $N$.) On the presumption that 
$K$ is a positive but decreasing function of $w$ in the neighbourhood of 
the critical value $w=0$, it can be seen that the ``magnetic'' side is 
characterised by $(C/N)^2<( 2\pi\, K_0)^2$ and the electric side by 
$(C/N)^2>( 2\pi\, K_0)^2$

In order to deal with both the ``magnetic'' and the ``electric'' cases
conjointly it is convenient to replace the mutually dual expressions 
given above for the mass function by yet another equivalent expression, 
namely
$$M=C\omega- 2\pi r\, L\, ,\eqno{(3.1)}$$
whose continuous variation leads directly to
$$ {\rm d}M=\omega\, {\rm d}C+2\pi\, K\, k\, {\rm d}N - 2 \pi\, T^{_{11}}
\,{\rm d}r\eqno{(3.2)}$$
with
$$ T^{_{11}}=L+ k^2 K=\tilde L+\omega^2 K\, ,\eqno{(3.3)}$$
which can be seen to be consistent with the expression in (2.2) for the 
relevant (space) component, with respect to the local stationary 
background frame, of the stress-energy tensor. The condition for 
mechanical equilibrium of the ring (namely ${\rm d}M=O$ for 
${\rm d}C= {\rm d}N=0$) is thus seen from (3.3) to be simply that the 
spatial stress-energy component should vanish, i.e.
$$ T^{_{11}}=0\, .\eqno{(3.4)}$$
This may be seen from (2.2) to be equivalent (in both the ``magnetic'' 
and ``electric'' cases) to the requirement that the string tension $T$ 
be related to its energy $U$ in the intrinsically preferred (corotating) 
frame by
$$T= r^2 {\mit\Omega}^2 U\, ,\eqno{(3.5)}$$
which can be seen to be interpetable as the condition that the rotation
velocity should coincide with the extrinsic characteristic velocity
$c_{_{\rm T}}$, the ``kink speed'', i.e.
 $$r^2 {\mit\Omega}^2=c_{_{\rm T}}^{\,2}\, ,\eqno{(3.6)}$$
where this speed $c_{_{\rm T}}$ of propagation (reltive to the locally
preferred frame) of transverse perturbations is given [2] by
$$c_{_{\rm T}}^{\,2}=T/U\, .\eqno{(3.7)}$$
We thus arrive at a theorem to the effect that the condition for 
equilibrium is that relatively backward moving perturbations should 
appear as standing waves, i.e. their angular velocity ${\mit\Omega}_-$ 
should vanish, the forward moving perturbations therefore having angular 
velocity, ${\mit\Omega}_+$ say, that in the non-relativistic limit would 
evidently have to be twice ${\mit\Omega}$, the exact result for the 
general case being 
$${\mit\Omega}_-=0\, ,\hskip 1 cm {\mit\Omega}_+= 2{\mit\Omega}/
(1+r^2{\mit\Omega}^2)\, .\eqno{(3.8)}$$

\bigskip\noindent
{\bf 4. Gyro-inertial and gyro-magnetic ratios}
\medskip

Subject to the foregoing equilibrium condition we are left with a family
of ring states determined by only two independent parameters, which may
be taken to be just the pair of conserved numbers $C$ and $N$.
Alternatively, in order to obtain all the relevant quantities in a more
explicit form, it may be more convenient (at least at a classical level
when one is dealing with value sufficiently large for the parametrisation 
to be considered as continuous) to take the independent parameters to
be the angular velocity ${\mit\Omega}$ and the state variable $w$,
since the latter immediately determines $\omega$ and $k$ (in magnitude
if not in sign) and hence also the phase velocity $\omega/k$, via the
mutually dual relations
$$\omega^2=-\tilde L/K\, ,\hskip 1 cm k^2=-L/K\, ,\eqno{(4.1)}$$
the value of the angular velocity variable ${\mit\Omega}$ then
determining both the sign of the phase velocity and the magnitude of the 
necessarily positive radius variable $r$ by the relation 
${\mit\Omega}\,r=\omega/k$ in the ``magnetic'' case for which the
phase velocity is subliminal, and ${\mit\Omega}\,r=k/\omega$ in the 
``electric'' case for which the phase velocity is superliminal

It is to be remarked that when the equilibrium condition (3.5) is
taken into account, the formula for the angular momentum of the ring 
can be converted into either of the dually related alternative forms
$$ J=2\pi r^3 U\,{\mit\Omega}=2\pi r^3 T\tilde {\mit\Omega}\, ,
\eqno{(4.2)}$$
while the mass of the ring can be expressed by the manifestly self
dual formula
$$M= ({\mit\Omega}+\tilde{\mit\Omega}) J\, .\eqno{(4.3)}$$
The angular momentum can also be expressed in manufestly self dual
form as
$$J=\pm 2\pi r^2\sqrt{UT}\, ,\eqno{(4.4)}$$
the analogous expression for the mass being
$$M=2\pi r(U+T)\, .\eqno{(4.5)}$$
Just as in the theory of pure vacuum black hole equilibrium states [10],
the solution is qualitatively determined (modulo an overall scale factor)
by the dimensionless ratio of mass $M$ to specific angular momentum
$a=J/M$, so analogously in the present context this same dimensionless
ratio $M^2/J$ (like $C/N$) fully determines the intrinsic state of the 
string as given by $w$ (leaving only the scale of the ring to be 
determined by an independent parameter such as ${\mit\Omega}$ or $r$
itself): we have 
$$M^2/J=\pm 2\pi(U+T)^2/\sqrt{UT}\, .\eqno{(4.6)}$$
(It can be seen that independently of $r$ this mass to specific angular
momentum ratio must always tend to the fixed value $8\pi\,T_0$ in the
low current limit as $w$ tends to zero, and for which both $T$ and $U$
tend to the common null state value $T_0$, which is consistent with 
finite charge for arbitrarily large vlues of $r$. Using the order of 
magnitude estimate $T_0\sim m^2$, where $m$ is the relevant symmetry 
breaking mass scale, one thus obtains a crude estimate for the numerical 
value of the ring mass as $M\sim|J|^{1/2}m$, the corresponding estimate 
for the radius being $r\sim |J|^{1/2} m^{-1}$. In the superconducting 
string applications one has in mind, it is not the grand unification mass 
scale that is contemplated, but less extreme ``charged ultra-massive 
particle'' scales that have been suggested in the literature [9]. The 
Compton wavelength $m^{-1}$ that is expected to characterise the string 
thickness will satisfy the requirement of being small compared with the 
estimated radius $r$ in the classical regime for which $|J|$ is large 
compared with unity, but it can be seen that the ``string'' description 
will break down in the quantum regime of small integer values of  $|J|$, 
for which one will otain a ``vorton'' more appropriately describable as a 
thick torus than as a thin ring.) 

Another scale independent function only of the intrinsic state of the 
string is the gyromagnetic factor, $g=2D/Qa$, which works out simply as
$$g=2MD/QJ=1+L/\tilde L\, ,\eqno{(4.7)}$$
whose translation into terms of $U$ and $T$ depends on whether we are 
dealing with a ``magnetic'' ring, in which case we shall have 
$g=1+c_{_{\rm T}}^{-2}$, or an ``electric'' ring , in which case we shall 
have $g=1+c_{_{\rm T}}^{\,2}$. It is of interest to notice that the 
intermediate case of a ``transluminal'' ring, i.e. in the limit as $U$ 
and $T$ both tend to $T_0$ so that $c_{_{\rm T}}$ tends to unity, the 
gyromagnetic ratio tends to the familiar value $g=2$ that characterises 
an electron in the simple Dirac theory, which is the same as has been 
observed [10] to apply to charged black hole equilibrium states. This 
value, $g=2$, has been rather inappropriately described as ``anomalous'' 
merely because it differs from the familiar value $g=1$ that holds for 
an ordinary massive charged rotating ring in the classical 
(nonrelativistic) limit. What is made apparent by the foregoing analysis 
is that in relativistic theory a stationary subluminally rotating string 
loop with a corotating electric charge, which is the case to which [1] 
the ``electric'' ring model applies (whether its underlying structure be
that of a vortex defect of the vacuum or something more mundane such as 
the twisted bunch of molecular chains that constitutes an ordinary 
rope), then (in view of the stability and causality requirements 
$0\leq T\leq U\ $ [2]) any  value in the range $1\leq g\leq 2$ is 
possible, the lower ``electrostatic'' limit being the classical value, 
and the upper ``transluminal'' limit being the Dirac value. On the other 
hand for a ``magnetic'' ring (corresponding to the case most commonly 
studied in recents work on cosmic string theory) the allowed range of 
values is $2\leq g\leq \infty$, the ``transluminal'' Dirac value being 
now the lower limit.

\bigskip\noindent
{\bf 5. A stability criterion}
\medskip

We conclude by remarking that the physical interest of the foregoing
result is dependent on the existence of parameter ranges for which
the stationary ring configurations are actually stable. We have been
considering states for which the derivatives of the mass function
(3.1) with respect to the radius at fixed values of the charge number
$C$ and winding number $N$ are given by
$${{ {\rm d}M}\over { {\rm d}r}}=0\, ,\hskip 1 cm  
{{ {\rm d}^2 M}\over{ {\rm d} r^2}}=
 {{ 2\pi U}\over {r}} {{ (3-c_{_{\rm T}}^{\,2})
c_{_{\rm T}}^{\,2}- (3 c_{_{\rm T}}^{\,2}-1) c_{_{\rm L}}^{\,2} }
\over {1-c_{_{\rm T}}^{\,2} c_{_{\rm L}}^{\,2}}} \eqno{(5.1)}$$
where $c_{_{\rm L}}^{\,2}$ is the longitudinal characteristic speed 
as given [2] by
$$c_{_{\rm T}}^{\,2}=- {\rm d}T/{\rm d}U\, .\eqno{(5.2)}$$
It may be conjectured that in addition to the local stability
conditions, the positivity of the second derivative in (5.1) is not
only necessary but also sufficient for the corresponding equilibrium 
state to be truly stable. This condition, i.e. the requirement that
the mass function has not just a critical value but actually a 
non-degenerate minimum can be seen to be expressible as the inequality
$$\left( {{c_{_{\rm T}} }\over {c_{_{\rm L}}}}\right)^2> 
{{ 3 c_{_{\rm T}}^{\,2}-1}\over {3 - c_{_{\rm T}}^{\,2}}}\, ,
\eqno{(5.3)}$$
whose validity is guaranteed by the positivity of the right hand side
whenever $c_{_{\rm T}}^{\,2}<{{_1}\over{^3}}$ (which will always be 
the case for classical string loops with $T\ll U$) but which may 
conceivably fail in certain cases (though not the most obvious ones)
in the relativistic context that is relevant to ``cosmic'' strings.

\bigskip\noindent
{\bf Acknowledgement}
\medskip

I should like to thank Nathalie Deruelle, Gary Gibbons, Patrick Peter,
Tsvi Piran, and peter Ruback, for helpful conversations.

\bigskip\noindent
{\bf References.}
\medskip
\parindent=0 cm

[1]  B. Carter, 
`Duality relation between charged elastic strings and superconducting
cosmic strings',
{\it Phys. Lett.} {\bf B224} (1989) 61.
\smallskip

[2]  B. Carter, 
`Stability and characteristic propagation  speeds in 
superconducting cosmic and other string Models', 
{\it Phys. Lett.} {\bf B228} (1989) 446.
\smallskip

[3] E. Witten, `Superconducting strings',
{\it Nucl. Phys.} {\bf B249} (1985) 557.
.\smallskip

[4] R.L. Davis, E.P.S. Shellard, 
`The physics of vortex superconductivity. 2', 
{\it Phys. Lett.} {\bf B209} (1988) 485.
\smallskip

[5] D.N. Spergel, T. Piran and J Goodman, 
`Dynamics of superconducting cosmic strings',
{\it Nucl. Phys.} {\bf B291} (1987) 847.
\smallskip

[6] A. Vilenkin, T. Vachaspati,
`Electromagnetic radiation from superconducting cosmic strings'
{\it Phys. Rev. Lett.} {\bf 58} (1987) 1041.
\smallskip

[7] E. Copeland, M. Hindmarsh, N. Turok, 
`Cosmic springs',
{\it Phys. Rev. Lett.} {\bf 56} (1987) 910.
\smallskip

[8] D. Haws, M. Hindmarsh, N. Turok, `Superconducting strings or 
springs?' 
{\it Phys. Lett.} {\bf B209} (1988) 255.
\smallskip

[9] A. De R\'ujula, S.L. Glashow and U. Sarid,
`Charged dark matter', CERN preprint TH5490 (2989).
{\it Nucl. Phys.} {\bf B333} (1990) 173. 
\smallskip

[10] B. Carter, `Mathematical foundations of the theory of 
relativistic stellar and black hole configurations' 
in {\it Gravitation in Astrophysics} (Carg\`ese, 1986), 
Nato A.S.I. {\bf  B156}, eds. B. Carter and J.B. Hartle
(Plenum, New York, 1987) 110.               

\end